\documentclass[runningheads,final]{llncs}
\usepackage{prelim2e}

\usepackage{ucs}
\usepackage[utf8x]{inputenc}
\usepackage[english]{babel}
\usepackage[spaces,hyphens]{url}
\usepackage{hyperref}

\usepackage[smaller]{acronym}
\usepackage{booktabs}
\usepackage[procnames]{listings}
  \lstMakeShortInline{\"}%
  \newcommand{\const}[1]{\textsc{#1}}%
  \newcommand{\id}[1]{\textit{#1}}%
  \newcommand{\proc}[1]{\textsc{#1}}%
  \newcommand{\nil}{\const{nil}}%
\usepackage{color}
\usepackage{float}%
  \floatstyle{ruled}%
  \newfloat{Algorithm}{htb}{lsa}%
\usepackage{graphicx}
\usepackage{longtable}
\usepackage{mdwtab}
\usepackage{multicol}
\usepackage{multirow}
\usepackage{paralist}
\usepackage{units}

\usepackage[reqno]{amsmath}
\usepackage{amsfonts} 
\usepackage{amssymb}
\newcommand{\Z}{\mathbb{Z}}
\newcommand{\inv}{\sp{-1}}

\usepackage{relsize}
\newcommand\ifmonospace{%
  \ifdim\fontdimen3\font=0pt%
}
\newcommand\Cpp{%
  \ifmonospace%
  C++%
  \else%
  C\kern-.0667em\raise.30ex\hbox{\relsize{-2}{++}}%
  \fi%
  \spacefactor1000%
}

\title{A novel parallel algorithm for Gaussian Elimination 
  of sparse unsymmetric matrices} 
\titlerunning{A parallel algorithm for Gaussian Elimination}
\author{Riccardo Murri}
\institute{%
    Grid Computing Competence Centre, \\
    Organisch-Chemisch Institut,
    University of Zürich, \\
    Winterthurerstrasse 190, CH-8006 Zürich, \\
    Switzerland. \\
    \email{riccardo.murri@gmail.com}
}
\date{Oct.~31, 2011}

\begin{document}

\PrerenderUnicode{üà}
\maketitle

\begin{abstract}
  We describe a new algorithm for Gaussian Elimination suitable for
  general (unsymmetric and possibly singular) sparse matrices of any
  entry type, which has a natural parallel and distributed-memory
  formulation but degrades gracefully to sequential execution.

  We present a sample \acs{MPI} implementation of a program computing
  the rank of a sparse integer matrix using the proposed algorithm.
  Some preliminary performance measurements are presented and
  discussed, and the performance of the algorithm is compared to
  corresponding state-of-the-art algorithms for floating-point and
  integer matrices.

\end{abstract}

\acrodef{BSP}{Bulk Synchronous Processes}
\acrodef{CPU}{Central Processing Unit}
\acrodef{GCC}{GNU C Compiler}
\acrodef{GE}{Gaussian Elimination}
\acrodef{GEPP}{Gaussian Elimination with Partial Pivoting}
\acrodef{IP}{Internet Protocol}
\acrodef{MPI}{Message-Passing Interface}
\acrodef{PU}{Processing Unit (see Section~\ref{sec:algo})}
\acrodef{SIMC}{Sparse Integer Matrices Collection (see \cite{SIMC})}
\acrodef{SLES}{SuSE Linux Enterprise Server}
\acrodef{SMS}{Sparse/Symbolic Matrix Storage}
\acrodef{TCP}{Transmission Control Protocol}
\acrodef{UFL}{university of Florida}
\acrodef{UZH}{University of Zurich}


\section{Introduction}
\label{sec:intro}

This paper presents a new algorithm for Gaussian Elimination,
initially developed for computing the rank of some homology matrices
with entries in the integer ring $\Z$.  It has a natural parallel
formulation in the message-passing paradigm and does not make use of
collective and blocking communication, but degrades gracefully to
sequential execution when run on a single compute node.

Gaussian Elimination algorithms with exact computations have been
analyzed in \cite{Dumas02computingthe}; the authors however concluded
that there was ---to that date--- no practical parallel algorithm for
computing the rank of sparse matrices, when exact computations are
wanted (e.g., over finite fields or integer arithmetic): well-known
Gaussian Elimination algorithms fail to be effective,
since, during elimination, entries in pivot position may become zero.

The ``Rheinfall'' algorithm presented here is based on the observation
that a sparse matrix can be put in a ``block echelon form'' with
minimal computational effort.  One can then run elimination on each
block of rows of the same length independently (i.e., in parallel);
the modified rows are then sent to other processors, which keeps the
matrix arranged in block echelon form at all times.  The procedure
terminates when all blocks have been reduced to a single row, i.e.,
the matrix has been put in row echelon form.

The ``Rheinfall'' algorithm is independent of matrix entry type, and
can be used for integer and floating-point matrices alike: numerical
stability is comparable with Gaussian Elimination with partial
pivoting (\acs{GEPP}).  However, some issues related to the computations with
inexact arithmetic have been identified in the experimental
evaluation, which suggest that ``Rheinfall'' is not a convenient
alternative to existing algorithms for floating-point matrices; see
Section~\ref{sec:lu-seq} for details.

Any Gaussian Elimination algorithm can be applied equally well to a
matrix column- or row-wise; here we take the row-oriented approach.

\section{Description of the ``Rheinfall'' Algorithm}
\label{sec:algo}

We shall first discuss the Gaussian Elimination algorithm for reducing
a matrix to row echelon form; practical applications like computing
matrix rank or linear system solving follow by simple modifications.

Let $A = (a_{ij} | i = 0, \ldots, n-1; j = 0, \ldots, m-1)$ be a $n
\times m$ matrix with entries in a ``sufficiently good'' ring $\Bbbk$ (a
field, a unique factorization domain or a principal ideal domain).
\begin{definition}
  Given a matrix $A$, let $z_i := \min \{j | a_{ij} \not= 0 \}$ be the
  column index of the first non-zero entry in the $i$-th row of $A$;
  for a null row, define $z_i := m$.  We say that the $i$-th row of
  $A$ \emph{starts} at column $z_i$.

  The matrix $A$ is in \emph{block echelon form} iff $z_i
  \geq z_{i-1}$ for all $i = 1, \ldots, n-1$.

  The matrix $A$ is in \emph{row echelon form} iff either $z_i >
  z_{i-1}$ or $z_i = m$.
\end{definition}
Every matrix can be put into \emph{block} echelon form by a
permutation of the rows.  For reducing the $n\times m$ matrix $A$ to
row echelon form, a ``master'' process starts $m$ Processing Units
$P[0]$, \ldots, $P[m-1]$, one for each matrix column: $P[c]$ handles
rows starting at column $c$.  Each Processing Unit (\acs{PU} for
short) runs the code in procedure \proc{ProcessingUnit} from
Algorithm~\ref{alg:echelon} concurrently with other \acsp{PU}; upon
reaching the \const{done} state, it returns its final output to the
``master'' process, which assembles the global result.
\begin{Algorithm}
  \caption{\label{alg:echelon}%
    Reduce a matrix to row echelon form by Gaussian
    Elimination. \emph{Left and top right:} Algorithm run by
    processing unit $P[c]$.  \emph{Bottom right:} Sketch of the
    ``master'' procedure.  Input to the algorithm is an $n \times m$
    matrix $A$, represented as a list of rows $r_i$. Row and column
    indices are $0$-based.}
  \begin{multicols}{2}
\begin{lstlisting}
def ProcessingUnit(c):
  u = |\const{nil}|
  Q = |empty list|
  output = |\const{nil}|
  state = |\const{running}|
  while state is |\const{running}|:
    |wait for messages to arrive|
    |append \const{Row} messages to $Q$|
    |select best pivot row $t$ from $Q$|
    if u is |\nil|:
      u = t
    else:
      if |$t$ has a better pivot than $u$|:
        |exchange $u$ and $t$|
    for |each row $r$| in $Q$: |\label{li:pu:loop1}|
        $r'$ = |\proc{eliminate}|(r,u) |\label{li:pu:elimination}|
        $c'$ = |first nonzero col.~of $r'$|
        |send $r'$ to $P[c']$|
        |delete $r$ from $Q$|
    if |received message \const{End}|:
      |wait for all sent messages to arrive|
      output = u                                         |\label{li:pu:result}|
      |send \const{End} to $P[c+1]$|
      state = |\const{done}|                             |\label{li:pu:loop2}|
  return output
\end{lstlisting}
\begin{lstlisting}
def Master(A):
  |start a \acs{PU} $P[c]$ for each column $c$ of $A$|  |\label{li:master:start}|
  for i in $\{0, \ldots, n-1\}$:                        |\label{li:master:read1}|
    c = |first nonzero column of $r_i$|
    |send $r_i$ to $P[c]$|                              |\label{li:master:read2}|
  |send \const{End} message to $P[0]$|                  |\label{li:master:core1}|
  |wait until $P[m-1]$ recv.~a \const{End} message|     |\label{li:master:core2}|
  result = |collect \id{output} from all \acs{PU}s|     |\label{li:master:result}|
  return result                                         |\label{li:master:end}|
\end{lstlisting}
  \end{multicols}
\end{Algorithm}

A Processing Unit can send messages to every
other \acs{PU}.  Messages can be of two sorts: \const{Row} messages
and \const{End} messages.  The payload of a \const{Row} message
received by $P[c]$ is a matrix row $r$, extending from column
$c$ to $m-1$; \const{End} messages carry no payload and just signal
the \acs{PU} to finalize computations and then stop execution.  In
order to guarantee that the \const{End} message is the last message
that a running \acs{PU} can receive, we make two assumptions on the
message exchange system:
\begin{inparaenum}[\itshape (1)]
\item that messages sent from one \acs{PU} to another arrive in the same
  order they were sent, and
\item that it is possible for a \acs{PU} to wait until all the messages it
  has sent have been delivered.
\end{inparaenum}
Both conditions are satisfied by \acs{MPI}-compliant message passing
systems.


The \proc{eliminate} function at line~\ref{li:pu:elimination} in
Algorithm~\ref{alg:echelon} returns a row $r' = \alpha r + \beta u$
choosing $\alpha, \beta \in \Bbbk$ so that $r'$ has a $0$ entry in
all columns $j \leq c$.  The actual definition of \proc{eliminate}
depends on the coefficient ring of $A$.
Note that $u[c] \not= 0$ by construction.

The ``master'' process runs the \proc{Master} procedure in
Algorithm~\ref{alg:echelon}. It is responsible for starting the $m$
independent Processing Units $P[0]$, \ldots, $P[m-1]$; feeding the
matrix data to the processing units at the beginning of the
computation; and sending the initial \const{End} message to \acs{PU} $P[0]$.
When the \const{End} message reaches the last Processing Unit, the
computation is done and the master collects the results.

Lines~\ref{li:master:read1}--\ref{li:master:read2} in \proc{Master}
are responsible for initially putting the input matrix~$A$ into block
echelon form; there is no separate reordering step. 
This is an invariant of the algorithm: by
exchanging rows among \acsp{PU} after every round of elimination is done,
the working matrix is kept in block echelon form at all times.

\begin{theorem}
  \label{thm:row-echelon-form}
  Algorithm~\ref{alg:echelon} reduces any given input matrix $A$ to
  row echelon form in finite time.
\end{theorem}
The simple proof is omitted for brevity.

\subsection{Variant: computation of matrix rank}
\label{sec:rank}

The Gaussian Elimination algorithm can be easily adapted to compute
the rank of a general (unsymmetric and possibly rectangular) sparse
matrix: one just needs to count the number of non-null rows of the row
echelon form.

Function \proc{ProcessingUnit} in Algorithm~\ref{alg:echelon}
is modified to return an integer number: the result shall be $1$ if at
least one row has been assigned to this \acs{PU} ($u \not= \nil$) and $0$
otherwise.

Procedure \proc{Master} performs a sum-reduce when collecting results:
replace line~\ref{li:master:result} with $\id{result} \gets$
sum-reduce of \id{output} from $P[c]$, for $c=1, \ldots, m$.

\subsection{Variant: LUP factorization}
\label{sec:lup}

We shall outline how Algorithm~\ref{alg:echelon} can be modified to
produce a variant of the familiar LUP factorization.  For the rest of
this section we assume that $A$ has coefficients in a field and is
square and full-rank.

It is useful to recast the Rheinfall algorithm in matrix
multiplication language, to highlight the small differences with the
usual LU factorization by Gaussian Elimination.  Let $\Pi^{}_0$ be the
permutation matrix that reorders rows of $A$ so that $\Pi^{}_0A$ is in
block echelon form; this is where Rheinfall's \acsp{PU} start their
work.  If we further assume that \acsp{PU} perform elimination and
only \emph{after that} they all perform communication at the same
time,\footnote{%
  As if using a \acs{BSP} model \cite{Valiant:1990} for
  computation/communication.  This assumption is not needed by
  ``Rheinfall'' (and is actually not the way it has been implemented)
  but does not affect correctness.  
}%
then we can write the $k$-th elimination step as multiplication by a
matrix $E_k$ (which is itself a product of elementary row operations
matrices), and the ensuing communication step as multiplication by a
permutation matrix $\Pi^{}_{k+1}$ which rearranges the rows again into
block echelon form (with the proviso that the $u$ row to be used for
elimination of other rows in the block comes first).  In other words,
after step $k$ the matrix $A$ has been transformed to $E_k
\Pi^{}_{k-1} \cdots E_0 \Pi^{}_0 A$.
\begin{theorem}
  Given a square full-rank matrix $A$, the Rheinfall algorithm outputs a
  factorization $\Pi A = L U$, where:
  \begin{itemize}
  \item $U = E_{n-1} \Pi^{}_{n-1} \cdots E_0 \Pi^{}_0 A$ is upper triangular;
  \item $\Pi = \Pi^{}_{n-1} \cdots \Pi^{}_0$ is a permutation matrix;
  \item $L = \Pi^{}_{n-1} \cdots \Pi^{}_1 \cdot E_0\inv \Pi\inv_1 E_1\inv
    \cdots \Pi\inv_{n-1} E\inv_{n-1}$ is lower unitriangular.
  \end{itemize}
\end{theorem}
The proof is omitted for brevity.

The modified algorithm works by exchanging triplets $(r, h, s)$ among
\acsp{PU}; every \acs{PU} stores one such triple $(u, i, l)$, and uses
$u$ as pivot row.  Each processing unit $P[c]$ receives a triple $(r,
h, s)$ and sends out $(r', h, s')$, where:
\begin{itemize}
\item The $r$ rows are initially the rows of $\Pi_0A$; they are modified by
  successive elimination steps as in Algorithm~\ref{alg:echelon}: $r'
  = r - \alpha u$ with $\alpha = r[c] / u[c]$.
\item $h$ is the row index at which $r$ originally appeared in $\Pi_0A$;
  it is never modified.
\item The $s$ rows start out as rows of the identity matrix: $s = e_h$
  initially.  Each time an elimination step is performed on $r$, the
  corresponding operation is performed on the $s$ row: $s' = s +
  \alpha l$.
\end{itemize}
When the \const{End} message reaches the last \acs{PU}, the \proc{Master}
procedure collects triplets $(u_c, i_c, l_c)$ from \acsp{PU} and constructs:
\begin{itemize}
\item the upper triangular matrix $U = (u_c)_{c=1,\ldots,n}$;
\item a permutation $\pi$ of the indices, mapping the initial
  row index $i_c$ into the final index $c$ (this corresponds to the
  $\Pi$ permutation matrix);
\item the lower triangular matrix $L$ by assembling the rows $l_c$
  after having permuted columns according to $\pi$.
\end{itemize}

\subsection{Pivoting}
\label{sec:stability}

A key observation in Rheinfall is that all rows assigned to a \acs{PU} start
at the same column.  This implies that pivoting is restricted to the
rows in a block, but also that each \acs{PU} may independently choose the
row it shall use for elimination.

A form of threshold pivoting can easily be implemented within these
constraints: assume that $A$ has floating-point entries and let $Q^+ =
Q \cup \{u\}$ be the block of rows worked on by Processing Unit $P[c]$
at a certain point in time (including the current pivot row $u$).  Let
$b = \max \left\{ |r[c]| : r \in Q^+ \right\}$; choose as pivot the
sparsest row $r$ in $Q^+$ such that $|r[c]| \geq \gamma \cdot b$,
where $\gamma \in [0,1]$ is the chosen threshold.
This guarantees that elements of $L$ are bounded by $\gamma\inv$.  

When $\gamma = 1$, threshold pivoting reduces to partial pivoting
(albeit restricted to block-scope), and one can repeat the error
analysis done in \cite[Section 3.4.6]{golub+van-loan} almost
verbatim.  The main difference with the usual column-scope partial
pivoting is that different pivot rows may be used at different times:
when a new row with a better pivoting entry arrives, it replaces the
old one.  This could result in the matrix growth factor being larger
than with \acs{GEPP}; only numerical experiments can tell how much
larger and whether this is an issue in actual practice.  However, no
such numerical experiments have been carried out in this preliminary
exploration of the Rheinfall algorithm.

Still, the major source of instability when using the Rheinfall
algorithm on matrices with floating-point entries is its sensitivity
to ``compare to zero'': after elimination has been performed on a row,
the eliminating \acs{PU} must determine the new starting column.  This
requires scanning the initial segment of the (modified) row to
determine the column where the first nonzero lies.  Changes in the
threshold $\epsilon>0$ under which a floating-point number is
considered zero can significantly alter the final outcome of Rheinfall
processing.

Stability is not a concern with exact arithmetic (e.g., integer
coefficients or finite fields): in this cases, leeway in choosing the
pivoting strategy is better exploited to reduce fill-in or avoid
entries growing too fast.  Experiments on which pivoting strategy
yields generally better results with exact arithmetic are still underway.

\section{Sample implementation}
\label{sec:impl}

A sample program has been written that implements matrix rank
computation and LU factorization with the variants of
Algorithm~\ref{alg:echelon} described before.  Source code is publicly
available from \url{http://code.google.com/p/rheinfall}.

Since there is only a limited degree of parallelism available on a
single computing node, processing units are not implemented as
separate continuously-running threads; rather, the
\texttt{ProcessingUnit} class provides a \texttt{step()} method, which
implements a single pass of the main loop in procedure
\proc{ProcessingUnit} (cf.\ lines~
\ref{li:pu:loop1}--\ref{li:pu:loop2} in Algorithm~\ref{alg:echelon}).
The main computation function consists of an inner loop that calls
each \acs{PU}'s \texttt{step()} in turn, until all \acsp{PU} have performed one
round of elimination. Incoming messages from other \acs{MPI} processes are then
received and dispatched to the destination \acs{PU}. 
This outer loop repeats until there are no more \acsp{PU} in
\const{Running} state.

When a \acs{PU} starts its \texttt{step()} procedure, it performs
elimination on all rows in its ``inbox'' $Q$ and immediately sends the
modified rows to other \acsp{PU} for processing.  Incoming messages
are only received at the end of the main inner loop, and dispatched to
the appropriate \acs{PU}.  Communication among \acsp{PU} residing in
the same \acs{MPI} process has virtually no cost: it is implemented by
simply adding a row to another \acs{PU}'s ``inbox''.  When
\acsp{PU} reside in different execution units, \texttt{MPI\_Issend} is
used: each \acs{PU} maintains a list of sent messages and checks at
the end of an elimination cycle which ones have been delivered and can
be removed.

\section{Sequential performance}
\label{sec:sequential}

The ``Rheinfall'' algorithm can of course be run on just one
processor: processing units execute a single \texttt{step()} pass
(corresponding to lines~\ref{li:pu:loop1}--\ref{li:pu:loop2} in
Algorithm~\ref{alg:echelon}), one after another; this continues until
the last \acs{PU} has switched to \const{Done} state.

\subsection{Integer performance}
\label{sec:rank-int-seq}

In order to get a broad picture of ``Rheinfall'' sequential
performance, the rank-computation program is being tested an all the
integer matrices in the \acs{SIMC} collection \cite{SIMC}.  A
selection of the results are shown in Table~\ref{tab:rf-vs-linbox},
comparing the performance of the sample Rheinfall implementation to
the integer \acs{GEPP} implementation provided by the free software
library \textsc{LinBox} \cite{linbox,linbox-site}.

Results in Table~\ref{tab:rf-vs-linbox} show great variability: the
speed of ``Rheinfall'' relative to \textsc{LinBox} changes by orders
of magnitude in one or the other direction.  The performance of both
algorithms varies significantly depending on the actual arrangement of
nonzeroes in the matrix being processed, with no apparent correlation
to simple matrix features like size, number of nonzeroes or fill
percentage.

Table~\ref{tab:transposes} shows the running time on the transposes of
the test matrices.  Both in \textsc{LinBox}'s \acs{GEPP} and in
``Rheinfall'', the computation times for a matrix and its transpose
could be as different as a few seconds versus several hours!  However,
the variability in Rheinfall is greater, and looks like it cannot be
explained by additional arithmetic work alone. More
investigation is needed to better understand how ``Rheinfall''
workload is determined by the matrix nonzero pattern.

\subsection{Floating-point performance}
\label{sec:lu-seq}

In order to assess the ``Rheinfall'' performance in floating-point
uses cases, the LU factorization program has been tested on a subset
of the test matrices used in~\cite{grigoridemmelli07}.  Results are
shown in Table~\ref{tab:rf-vs-superlu}, comparing the Mflop/s attained
by the ``Rheinfall'' sample implementation with the performance of
\textsc{SuperLU 4.2} on the same platform.

The most likely cause for the huge gap in performance between
``Rheinfall'' and \textsc{SuperLU} lies in the strict row-orientation
of ``Rheinfall'': \textsc{SuperLU} uses block-level operations,
whereas Rheinfall only operates on rows one by one.  
However, row orientation is a defining characteristics of the
``Rheinfall'' algorithm (as opposed to a feature of its
implementation) and cannot be circumvented.  Counting also the
``compare to zero'' issue outlined in Section~\ref{sec:stability}, one
must conclude that ``Rheinfall'' is generally not suited for inexact
computation.

\section{Parallel performance and scalability}
\label{sec:parallel}

The ``Rheinfall'' algorithm does not impose any particular scheme for
mapping \acsp{PU} to execution units.  A column-cyclic scheme has been
currently implemented.

Let $p$ be the number of \acs{MPI} processes (ranks) available, and
$m$ be the total number of columns in the input matrix $A$.  The input
matrix is divided into vertical stripes, each comprised of $w$
adjacent columns. Stripes are assigned to \acs{MPI} ranks in a cyclic
fashion: \acs{MPI} process $k$ (with $0 \leq k < p$) hosts the $k$-th,
$(k+p)$-th, $(k+2p)$-th, etc.\ stripe; in other words, it owns
processing units $P[w \cdot (k + a \cdot p) + b]$ where $a = 0, 1,
\ldots$ and $0 \leq b < w$.

\subsection{Experimental results}
\label{sec:scalability}

\begin{figure*}[p]
  \begin{center}
    \includegraphics[width=0.49\linewidth]{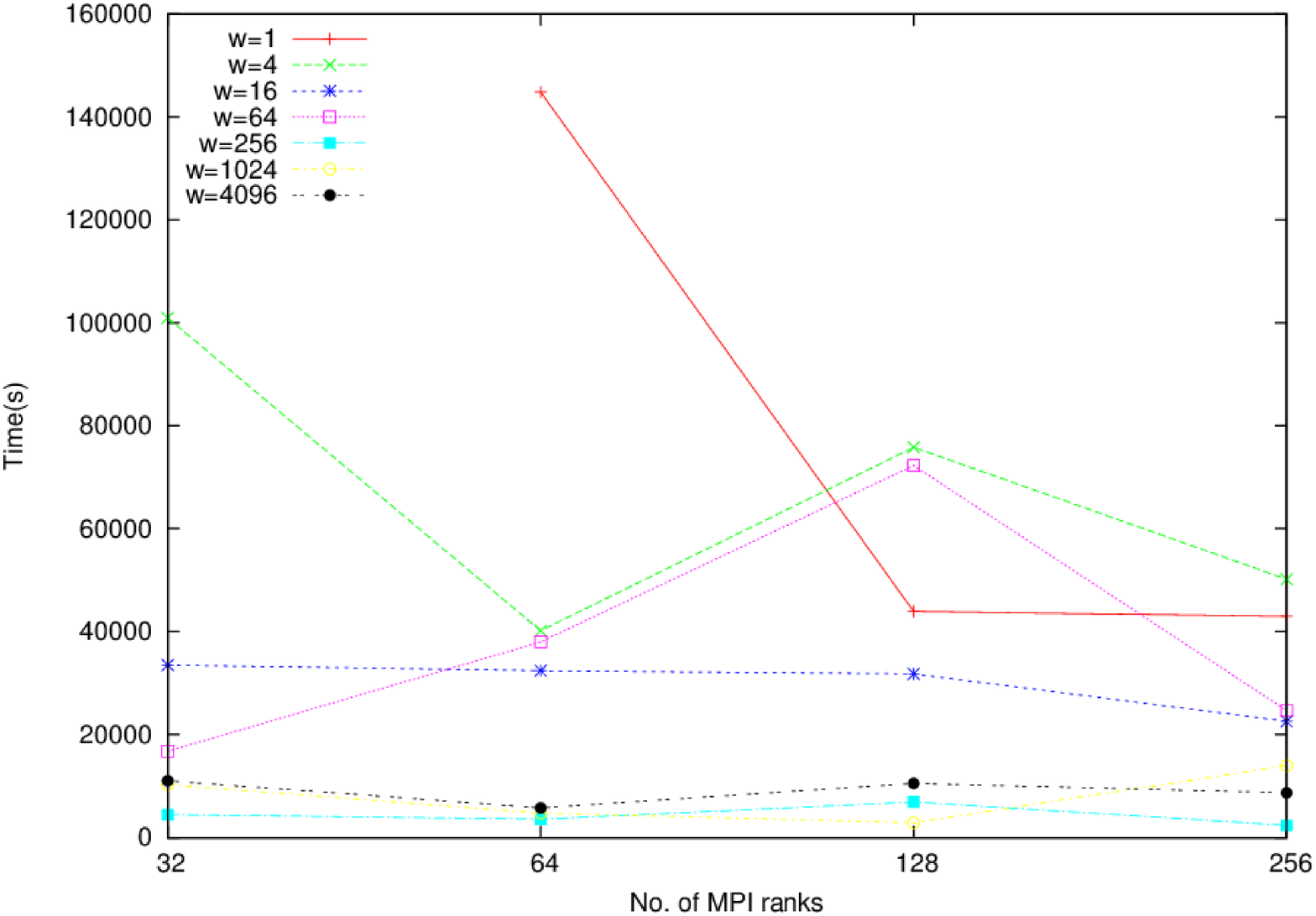}
    \includegraphics[width=0.49\linewidth]{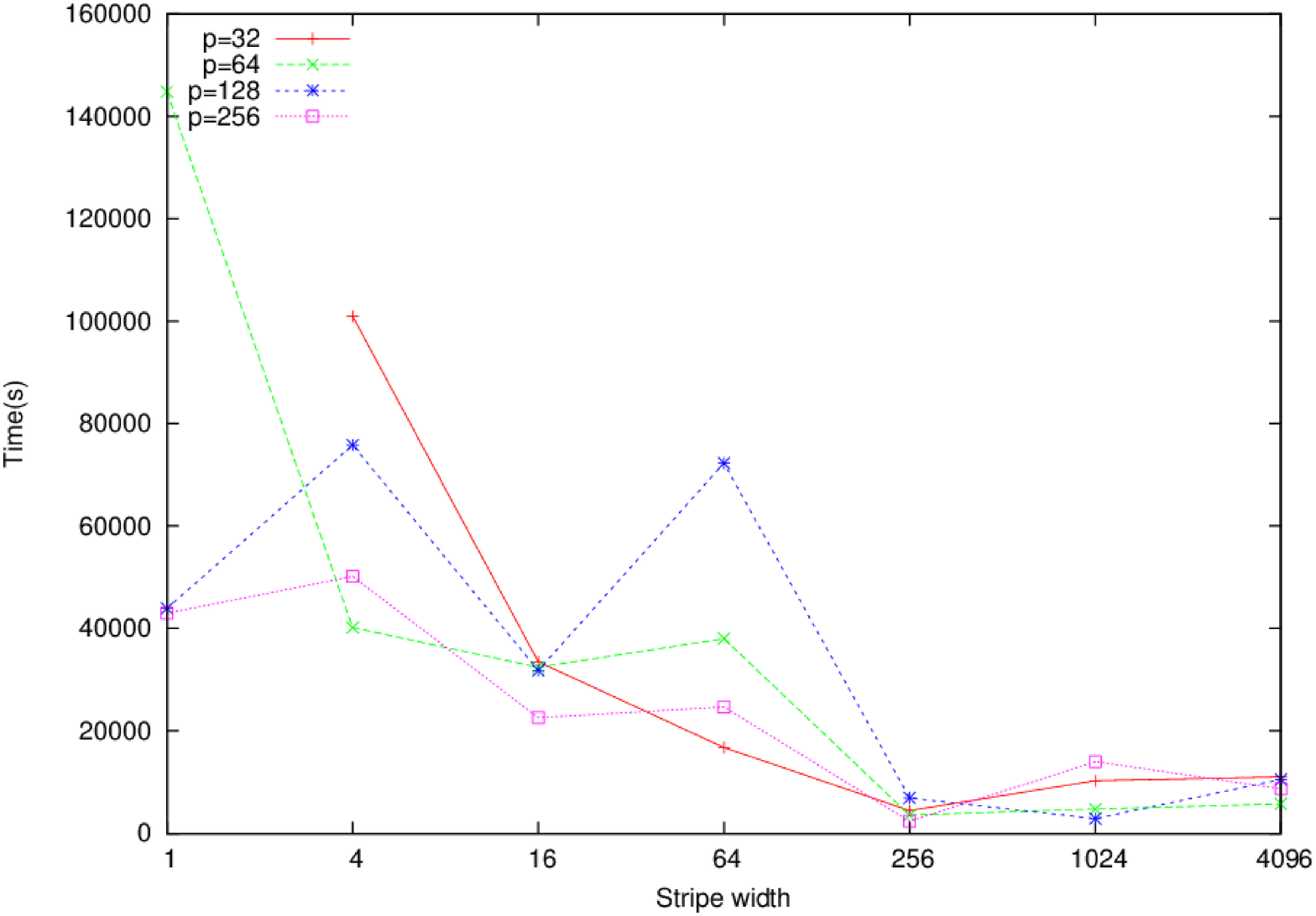}
  \end{center}
  \caption{\label{fig:times}%
    \emph{Left:} Plot of the running time (in seconds, $y$-axis) of the
    sample Rheinfall implementation on the matrix M0,6-D10, versus the
    number $p$ of \acs{MPI} ranks ($x$-axis). 
    \emph{Right:} Plot of the running time (in seconds, $y$-axis) of
    the sample Rheinfall implementation on the matrix M0,6-D10, versus
    stripe width $w$ ($x$-axis).
  }%
\end{figure*}

In order to assess the parallel performance and scalability of the
sample ``Rheinfall'' implementation, the rank-computation program has
been run on the matrix \texttt{M0,6-D10} (from the \textsc{Mgn} group
of \acs{SIMC} \cite{SIMC}; see Table~\ref{tab:rf-vs-linbox} for
details).  The program has been run with a varying number of \acs{MPI}
ranks and different values of the stripe width parameter $w$: see
Figure~\ref{fig:times}.

The plots in Figure~\ref{fig:times} show that running time generally
decreases with higher $w$ and larger number $p$ of \acs{MPI} ranks
allocated to the computation, albeit not regularly.  This is
particularly evident in the plot of running time versus stripe width
(Figure~\ref{fig:times}, right), which shows an alternation of
performance increases and decreases.  A more detailed investigation is
needed to explain this behavior; we can only present here some
working hypotheses.

The $w$ parameter influences communication in two different ways. On
the one hand, there is a lower bound $O(m/w)$ on the time required to
pass the \const{End} message from $P[0]$ to $P[m]$.  Indeed, since the
\const{End} message is always sent from one \acs{PU} to the next one,
then we only need to send one \const{End} message per stripe over the
network. This could explain why running time is almost the same for
$p=128$ and $p=256$ when $w=1$: it is dominated by the time taken to
pass the \const{End} message along.

On the other hand, \acs{MPI} messages are collected after each
processing unit residing on a \acs{MPI} rank has performed a round of
elimination; this means that a single \acs{PU} can slow down the
entire \acs{MPI} rank if it gets many elimination operations to
perform.  The percentage of running time spent executing \acs{MPI}
calls has been collected using the \texttt{mpiP} tool \cite{mpiP-web};
a selection of relevant data is available in Table~\ref{tab:mpiP}.
The three call sites for which data is presented measure three
different aspects of communication and workload balance:
\begin{itemize}
\item The \texttt{MPI\_Recv} figures measure the time spent in actual
  row data communication (the sending part uses \texttt{MPI\_Issend}
  which returns immediately).
\item The \texttt{MPI\_Iprobe} calls are all done after all \acsp{PU}
  have performed one round of elimination: thus they measure the time
  a given \acs{MPI} rank has to wait for data to arrive.
\item The \texttt{MPI\_Barrier} is only entered after all
  \acsp{PU} residing on a given \acs{MPI} rank have finished their job;
  it is thus a measure of workload imbalance.
\end{itemize}

Now, processing units corresponding to higher column indices naturally
have more work to do, since they get the rows at the end of the
elimination chain, which have accumulated fill-in.  Because of the way
\acsp{PU} are distributed to \acs{MPI} ranks, a larger $w$ means that
the last \acs{MPI} rank gets more \acsp{PU} of the final segment: the
elimination work is thus more imbalanced.  This is indeed reflected in
the profile data of Table~\ref{tab:mpiP}: one can see that the maximum
time spent in the final \texttt{MPI\_Barrier} increases with $w$ and
the number $p$ of \acs{MPI} ranks, and can even become 99\% of the
time for some ranks when $p=256$ and $w=4096$.

However, a larger $w$ speeds up delivery of \const{Row} messages from
$P[c]$ to $P[c']$ iff $(c' - c) / w \equiv 0 (\textrm{mod~} p)$. 
Whether this is beneficial is highly dependent on the structure
of the input matrix: internal regularities of the input
data may result on elimination work being concentrated on the same
\acs{MPI} rank, thus slowing down the whole program.  Indeed, the
large percentages of time spent in \texttt{MPI\_Iprobe} for some
values of $p$ and $w$ show that the matrix nonzero pattern plays a big
role in determining computation and communication in Rheinfall.
Static analysis of the entry distribution could help determine an
assignment of \acsp{PU} to \acs{MPI} ranks that keeps the work more
balanced.

\section{Conclusions and future work}
\label{sec:future}

The ``Rheinfall'' algorithm is basically a different way of arranging
the operations of classical Gaussian Elimination, with a naturally
parallel and distributed-memory formulation.  It retains some
important features from the sequential Gaussian Elimination; namely,
it can be applied to general sparse matrices, and is independent of
matrix entry type.  Pivoting can be done in Rheinfall with strategies
similar to those used for \acs{GEPP}; however, Rheinfall is not
equally suited for exact and inexact arithmetic.

Poor performance when compared to state-of-the-art algorithms and some
inherent instability due to the dependency on detection of nonzero
entries suggest that ``Rheinfall'' is not a convenient alternative for
floating-point computations.

For exact arithmetic (e.g., integers), the situation is quite the
opposite: up to our knowledge, ``Rheinfall'' is the first practical
distributed-memory Gaussian Elimination algorithm capable of exact
computations.  In addition, it is competitive with existing
implementations also when running sequentially.

The distributed-memory formulation of ``Rheinfall'' can easily be
mapped on the \acs{MPI} model for parallel computations.  An issue
arises on how to map Rheinfall's Processing Units to actual \acs{MPI}
execution units; the simple column-cyclic distribution discussed in
this paper was found experimentally to have poor workload
balance. Since the workload distribution and the communication graph
are both determined by the matrix nonzero pattern, a promising future
direction could be to investigate the use of graph-based partitioning
to determine the distribution of \acsp{PU} to \acs{MPI} ranks.

\section*{Acknowledgments}
\label{sec:ack}

The author gratefully acknowledges support from the University of
Zurich and especially from Prof.\ K. K.~Baldridge, for the use of the
computational resources that were used in developing and testing the
implementation of the Rheinfall algorithm.  Special thanks go to
Professors\ J.~Hall and O.~Schenk, and Dr.\ D.~Fiorenza for fruitful
discussions and suggestions.  Corrections and remarks from the PPAM
reviewers greatly helped the enhancement of this paper from its
initial state.

\bibliographystyle{splncs03}
\bibliography{linalg}

\begin{table*}[p]
  \begin{center}
    \begin{tabular}{l@{\extracolsep{1em}}rrrrrr}
      \toprule
      \textsc{Matrix} &    rows  &  columns  &  nonzero  &  fill\%  &        Rheinfall &    LinBox       \\%
      \midrule                                            
      M0,6-D8         &  862290  &  1395840  &  8498160  &  0.0007  & \textbf{23.81}   &        36180.55 \\%
      M0,6-D10        &  616320  &  1274688  &  5201280  &  0.0007  &        {23378.86}&\textbf{13879.62}\\%
      olivermatrix.2  &   78661  &   737004  &  1494559  &  0.0026  & \textbf{2.68}    &          115.76 \\%
      Trec14          &    3159  &    15905  &  2872265  &  5.7166  & \textbf{116.86}  &          136.56 \\%
      GL7d24          &   21074  &   105054  &   593892  &  0.0268  &  95.42           &   \textbf{61.14}\\%
      IG5-18          &   47894  &    41550  &  1790490  &  0.0900  & 1322.63          &   \textbf{45.95}\\%
      \bottomrule
    \end{tabular}
  \end{center}
  \caption{\label{tab:rf-vs-linbox}%
    \acs{CPU} times (in seconds) for computing the matrix rank of
    selected integer matrices; boldface font marks the best result in
    each row.  The ``Rheinfall'' column reports times for the sample
    \Cpp{} implementation. The ``LinBox'' column reports times for the
    \acs{GEPP} implementation in \textsc{LinBox} version 1.1.7.
    The programs were run on the \acs{UZH}
    ``Schroedinger'' cluster, equipped with Intel Xeon X5560 \acsp{CPU} @
    2.8GHz and running 64-bit \acs{SLES} 11.1 Linux; codes were compiled
    with \acs{GCC} 4.5.0 with options \texttt{-O3~-march=native}.
  }%
\end{table*}

\begin{table*}[p]
  \begin{center}%
\begin{tabular}{l@{\extracolsep{1em}}rrrr}
  \toprule
 {\sc Matrix}& Rheinfall \emph{(T)} &  Rheinfall &LinBox  \emph{(T)}&   LinBox       \\  
 \midrule                                                                 
 M0,6-D8         & \emph{No~mem.}  & {\bf 23.81}&        50479.54  &      36180.55  \\  
 M0,6-D10        & \textbf{37.61}  &  {23378.86}&        26191.36  &      13879.62  \\  
 olivermatrix.2  & \textbf{0.72}   &  {2.68}    &          833.49  &        115.76  \\  
 Trec14          & \emph{No~mem.}  &   116.86   &  \textbf{43.85}  &        136.56  \\
 GL7d24          & \textbf{4.81}   &    95.42   &          108.63  &         61.14  \\  
 IG5-18          &     12303.41    &  1322.63   &          787.05  & \textbf{45.95}  \\
 \bottomrule  
\end{tabular}
\end{center}                                  
\caption{\label{tab:transposes}%
  \acs{CPU} times (in seconds) for computing the matrix rank of selected integer
  matrices and their transposes; boldface font marks the best result in
    each row.  The table compares running times of
  the Rheinfall/\Cpp{} and \acs{GEPP} LinBox 1.1.7 codes.  The
  columns marked with \emph{(T)} report \acs{CPU} times used for the
  transposed matrix. Computation of the transposes of matrices ``M0,6-D8'' and ``Trec14'' exceeded
  the available $\unit[24]{GB}$ of RAM.
  Hardware, compilation flags and running conditions are as in
  Table~\ref{tab:rf-vs-linbox}, which see also for matrix size and other
  characteristics.
}%
\end{table*}

\begin{table*}[p]
  \begin{center}
    \begin{tabular}{l@{\extracolsep{1em}}rrrlr}
      \toprule
      \textsc{Matrix}&     $N$  &  nonzero  &  fill\% & Rheinfall  &  SuperLU          \\
      \midrule
      bbmat          &   38744  &  1771722  &  0.118  &      83.37 &  \textbf{1756.84} \\
      g7jac200sc     &   59310  &   837936  &  0.023  &      87.69 &  \textbf{1722.28} \\
      lhr71c         &   70304  &  1528092  &  0.030  &{\emph{No mem.}}
                                                                   &  \textbf{ 926.34} \\
      mark3jac140sc  &   64089  &   399735  &  0.009  &      92.67 &  \textbf{1459.39} \\
      torso1         &  116158  &  8516500  &  0.063  &      97.01 &  \textbf{1894.19} \\
      twotone        &  120750  &  1224224  &  0.008  &      91.62 &  \textbf{1155.53} \\
      \bottomrule
    \end{tabular}
  \end{center}
  \caption{\label{tab:rf-vs-superlu}%
    Average Mflop/s attained in running LU factorization of
    square $N \times N$ matrices; boldface font marks the best result in
    each row. The table compares the performance
    of the sample Rheinfall/\Cpp{} LU factorization with
    \textsc{SuperLU}~4.2. The test
    matrices are a subset of those used in \cite{grigoridemmelli07}.
    See Table~\ref{tab:rf-vs-linbox} for hardware characteristics.
  }
\end{table*}

\begin{table*}[p]
  \centering
  \begin{tabular}{c@{\extracolsep{1ex}}c|ccc|cc|cc|cc}
    \toprule
    \multirow{2}{*}{$p$} & \multirow{2}{*}{$w$} 
                & \multicolumn{3}{c|}{MPI Total\%} 
                                                   &\multicolumn{2}{c|}{\sc MPI\_Recv\%}
                                                                   &\multicolumn{2}{c|}{\sc MPI\_Iprobe\%}
                                                                                   & \multicolumn{2}{c}{\sc MPI\_Barrier\%} \\
        &       & Avg.$\pm \sigma$  &  Max. & Min.  & Avg.  & Max.  & Avg.  & Max.  & Avg.  & Max. \\
    \midrule
    16  & 16    & $18.79 \pm 0.32$  & 19.61 & 18.37 & 79.96 & 81.51 & 3.13  & 3.29  & 0.00 & 0.00 \\
    32  & 16    & $11.54 \pm 0.53$  & 12.87 & 10.93 & 4.41  & 71.97 & 14.19 & 14.97 & 0.00 & 0.00 \\
    64  & 16    & $12.25 \pm 0.20$  & 12.78 & 11.77 & 1.12  & 55.94 & 34.57 & 36.13 & 0.00 & 0.00 \\
    128 & 16    & $20.51 \pm 0.55$  & 23.87 & 19.33 & 31.62 & 35.06 & 61.08 & 64.17 & 0.00 & 0.00 \\
    \hline
    16  & 256   & $26.77 \pm 1.79$  & 29.50 & 23.29 & 89.69 & 92.31 & 1.27  & 1.50  & 0.00 & 0.20 \\
    32  & 256   & $10.17 \pm 1.34$  & 13.57 & 7.83  & 78.51 & 85.33 & 13.10 & 18.08 & 0.00 & 0.01 \\
    64  & 256   & $16.55 \pm 2.16$  & 22.15 & 11.74 & 61.46 & 73.13 & 27.20 & 43.73 & 0.00 & 0.67 \\
    128 & 256   & $15.43 \pm 0.64$  & 18.65 & 14.35 & 6.15  & 10.97 & 90.98 & 95.27 & 0.00 & 0.02 \\
    256 & 256   & $38.92 \pm 1.94$  & 43.24 & 32.99 & 6.12  & 14.20 & 89.38 & 97.41 & 0.00 & 0.60 \\
    \hline
    16  & 4096  & $9.08 \pm 1.62$   & 13.81 & 7.22  & 50.57 & 66.97 & 7.52  & 10.35 & 0.00  & 0.19  \\
    32  & 4096  & $6.53 \pm 1.97$   & 12.33 & 3.71  & 36.80 & 58.48 & 28.30 & 51.23 & 1.51  & 3.71  \\
    64  & 4096  & $6.81 \pm 1.52$   & 12.01 & 4.53  & 8.73  & 30.15 & 72.13 & 93.92 & 10.88 & 26.93 \\
    128 & 4096  & $16.73 \pm 7.80$  & 44.72 & 8.59  & 5.12  & 21.82 & 46.82 & 92.24 & 43.69 & 86.65 \\
    256 & 4096  & $45.78 \pm 28.32$ & 88.22 & 9.91  & 0.00  & 9.18  & 11.94 & 96.68 & 86.09 & 98.63 \\
    \bottomrule
  \end{tabular}
  \caption{\label{tab:mpiP}%
    Percentage of running time spent in \acs{MPI} communication for the
    sample Rheinfall/C implementation on the matrix \textsc{M0,6-D10},
    with varying number of \acs{MPI} ranks and stripe width parameter $w$.
    Columns \texttt{MPI\_Recv}, \texttt{MPI\_Iprobe} and
    \texttt{MPI\_Barrier} report on the percentage of \acs{MPI} time spent
    spent servicing these calls; in these cases, the minimum is always
    very close to zero hence it is omitted from the table.  Tests were
    executed on the \acs{UZH} cluster ``Schroedinger''; see
    Table~\ref{tab:rf-vs-linbox} for hardware details. The \acs{MPI} layer
    was provided by OpenMPI version 1.4.3, using the \acs{TCP}/\acs{IP}
    transport.
  }%
\end{table*}

\end{document}
